\documentclass[12pt]{iopart}
\usepackage{amssymb}
\usepackage[dvips]{graphicx}
\usepackage{enumerate}
\usepackage{epsfig}
\usepackage{subfigure}
\usepackage{xcolor}
\usepackage[T1]{fontenc}
\usepackage{fullpage}
\usepackage{amsthm,amsfonts,amssymb,amscd,mathrsfs,xspace,framed}
\usepackage{mathrsfs,amsmath}
\usepackage{color}
\usepackage{setspace}
\usepackage{url}
\usepackage{wrapfig}
\usepackage{tikz}
\usepackage{enumitem}
\usepackage{bm}
\usepackage{comment}
\usepackage{txfonts}
\usepackage{array}
\usepackage{cancel}
\usepackage{braket}
\usepackage{mathtools}

\usepackage[hyperfootnotes=false]{hyperref}
\usepackage[acronym]{glossaries}
\usepackage[english]{babel}
\usepackage{url}
\usepackage{bm}
\usepackage{hyperref}
\definecolor{darkblue}{rgb}{0,0,0.5}
\hypersetup{
colorlinks=true,
linkcolor=black,
filecolor=blue,
citecolor=darkblue,  
urlcolor=black,
}

\usepackage[style=numeric,sorting=none]{biblatex}
\newcommand{\BH}[1]{{{\textcolor{black}{#1}}}}

\begin{document}

\title[Continuous-variable quantum repeaters]{Continuous-variable quantum repeaters based on bosonic error-correction and teleportation: architecture and applications}

\author{Bo-Han Wu$^{1}$, Zheshen Zhang$^{2,3,4}$,
Quntao Zhuang$^{3,4}$}
\address{$^1$ Department of Physics, University of Arizona, Tucson, Arizona 85721, USA}
\address{$^2$ Department of Materials Science and Engineering, University of Arizona, Tucson, Arizona 85721, USA}
\address{$^3$ Department of Electrical and Computer Engineering, University of Arizona, Tucson, Arizona 85721, USA}
\address{$^4$ J. C. Wyant College of Optical Sciences, University of Arizona, Tucson, Arizona 85721, USA}
\ead{gowubohan@email.arizona.edu}

\vspace{10pt}
\begin{indented}
\item[]Sep 2021
\end{indented}

\begin{abstract}
Quantum repeaters are essential ingredients for quantum networks that link distant quantum modules such as quantum computers and sensors. Motivated by distributed quantum computing and communication, quantum repeaters that relay discrete-variable quantum information have been extensively studied; while continuous-variable (CV) quantum information underpins a variety of quantum sensing and communication application, a quantum-repeater architecture for genuine CV quantum information remains largely unexplored. This paper reports a CV quantum-repeater architecture based on CV quantum teleportation assisted by the Gottesman-Kitaev-Preskill (GKP) code to significantly suppress the physical noise. The designed CV quantum-repeater architecture is shown to significantly improve the performance of CV quantum key distribution, entanglement-assisted communication, and target detection based on quantum illumination, as three representative use cases for quantum communication and sensing.
\end{abstract}

\date{\today}

\section{Introduction: file preparation and submission}
Quantum networks~[1-6] not only offer unconditional security in private-key distributions~[7-10], but also enable the establishment of entanglement across multiple parties to endow quantum-enhanced capabilities. Photons are ideal information carriers for long-haul quantum communications by virtue of their robustness against environmental noise, but they are susceptible to loss because, unlike classical information, quantum information cannot be regenerated by amplifiers due to the quantum no-cloning theorem~[11,12]. Such a restriction places a fundamental rate-loss trade-off between entanglement-distribution rate and transmission distance, which, in terms of the distribution of bipartite entanglement, was formulated as the Pirandola--Laurenza--Ottaviani--Banchi (PLOB) bound~[13] and has been subsequently generalized to end-to-end capacity of a general quantum network~[14]. 

To circumvent the rate-loss trade-off, a long-distance quantum link is divided into shorter and less lossy links via introducing intermediate quantum repeater (QR) nodes~[15-21]. Based on the processing power at each node, QRs are categorized into three generations (see Refs.~[15,22,23], Refs.~[17,24] and Refs.~[25,26]). The mainstream QR architectures have been dedicated to the long-distance distribution of discrete-variable (DV) quantum states~[17,27-29], i.e., qubits, to link quantum computers, in analogy to sharing digital information among classical computers. On the other hand, continuous-variable (CV) quantum states, akin to analog information, underpins a variety of quantum-enhanced sensing and communication capabilities including entangled sensor networks~[30-37], physical-layer quantum data classification~[38,39], quantum-illumination (QI) target detection~[40-43] and ranging~[44], and entanglement-assisted (EA) communication~[45-53]. Apart from a handful of investigations for a few specific use cases~[54,55], the QR architecture for CV quantum states remains largely unexplored.

\begin{figure}
	{\centering\includegraphics[width=1\linewidth]{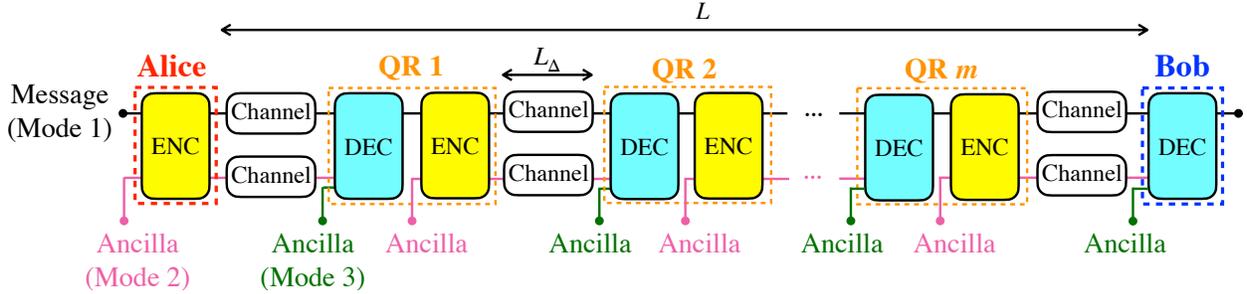}}
	\caption{Scheme of $m$-relay repeaters based on CV error-correction protocol. ENC: encoding. DEC: decoding. $L$ is the physical distance between Alice and Bob. $L_{\Delta}$ is the inter-repeater spacing.}
	\label{fig:general}
\end{figure}

Quantum error correction (QEC) is an essential ingredient for QRs to reliably relay quantum information. QEC for qubits has been well established to support the development of fault-tolerant quantum computing~[56,57]. QEC for QRs, however, requires an additional framework to account for the infinite dimensional Hilbert space that photons reside in. In this regard, bosonic QEC~[58] has emerged as a powerful paradigm to protect quantum information carried on photons. To date, multiple single-mode bosonic codes, including the binomial code~[59,60], Schr\"odinger-cat-state codes~[61-65], and \emph{Gottesman-Kitaev-Perskill} (GKP) codes~[66-70], have been proposed and experimentally produced \BH{in the platforms of trapped ion and superconducting qubit}~[71-75]. Most bosonic codes have been designed to protect qubits by encoding them into bosonic modes. The more recent works of Rozp\c{e}dek \emph{et al.}~[76] and Fukui \emph{et al.}~[77,78] introduced the \BH{optical} GKP-formed qubit codes into the QR architecture to transmit qubits, but a QR based on bosonic QEC to transmit CV quantum information, which will significantly benefit a wide range of quantum-enhanced applications, remains elusive. \BH{ While generating optical GKP states in the experiment is still challenging, recently, a few theoretical works have been proposed on generating optical GKP state probabilistically~[79-81] or deterministically~[82].}

This article proposes a CV QR architecture based on the recently developed GKP-assisted QEC~[67,83] combined with CV quantum teleportation~[84,85] and is organized as follows: Sec.~\ref{sec:QRA} provides an overview on the QR architecture; Sec.~\ref{sec:background} introduces the GKP-assisted QEC to the QR architecture. Finally, in Sec.~\ref{sec:App}, the QEC QR is shown to boost the performance of EA communication, target detection based on QI, and CV quantum key distribution (QKD).

\section{Quantum-repeater architecture with bosonic quantum error correction}
\label{sec:QRA}

Fig.~\ref{fig:general} illustrates the architecture for our CV QR based on the bosonic QEC code~[67]. Consider a quantum link comprising $m$ QR nodes. At the source, Alice performs an \emph{encoding} operation on the message mode and an ancilla mode and then transmits both modes to the first QR node through a quantum channel. The QR node performs a \emph{decoding} operation on both received modes to correct the accumulated errors incurred by the channel. Afterwards, \emph{encoding} operations are operated on the error-corrected message mode and an additional ancilla mode; the two modes are subsequently transmitted to the next QR node for decoding and encoding, until the message mode is finally decoded at Bob's terminal.

Note that here the quantum channels not only model the transmission via fiber quantum links, but also takes into account some pre- and post-processing that enhances the quantum information transmission. Each fiber link between two nodes can be modeled as a bosonic pure-loss channel with the transmissivity $\eta=10^{-\gamma L_{\Delta}/10}$, where $L_{\Delta}$ is the physical distance between the two nodes, with an attenuation factor $\gamma=0.2$~decibels per kilometer. With additional pre- and post-processing, we convert the pure-loss link into two types of quantum channels, the amplified one-way channel (Sec.~\ref{sec:amp}) and the quantum teleportation channel (Sec.~\ref{sec:tele}). The effect of transmitting the message and ancilla modes through the amplified one-way or quantum teleportation channel is equivalent to adding to their quadratures some additive noises of variance $\sigma^2_A$ or $\sigma^2_T$, instead of the original pure-loss.

\begin{figure}[t]
	{\centering\includegraphics[width=0.5\linewidth]{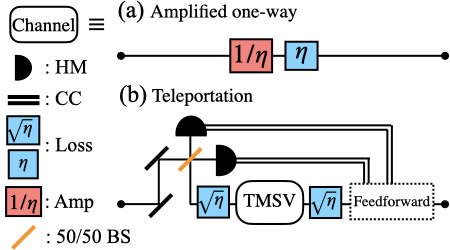}}
	\centering
	\caption{The scheme of (a) amplified one-way channel, and (b) teleportation channel. CC: classical communication. HM: homodyne measurement. Amp: amplification. \BH{BS: beamsplitter.}\label{fig:protocol}}
	{\centering\includegraphics[width=0.47\linewidth]{Additivenoise.png}}
	\caption{Variances of the additive noise for the protocols of amplified one-way and teleportation channels. Amp: amplification. Tele: teleportation.  \label{fig:additivenoise}}
\end{figure}

\subsection{Amplified one-way channel} 
\label{sec:amp} 
Sketched in Fig.~\ref{fig:protocol}(a), the amplified one-way channel introduced in the QR architecture studied by Fukui \emph{et al.}~[77] applies a phase-insensitive amplifier of gain $1/\eta$ before the pure-loss channel of transmissivity $\eta$ induced by the fiber transmission. The variance of additive noise of the amplified one-way channel is derived to be
\begin{equation}
    \sigma^2_{A}=1-\eta,
\label{eq:p1}
\end{equation}
i.e., $\braket{\hat{q}^2}_{\text{vac}}=\braket{\hat{p}^2}_{\text{vac}}=1/2$. Because both the channel loss and the amplification add noise, the performance of QEC is limited. To overcome the drawback of the amplified one-way channel, we introduce the quantum teleportation channel below.

\subsection{Quantum teleportation channel}
\label{sec:tele}
CV quantum teleportation transmits CV quantum states from the source to the destination with only local operations, classical communication (LOCC), and shared CV entangled states. To implement a CV quantum teleportation channel in the CV QR architecture, a two-mode squeezed vacuum (TMSV) source placed in the middle of QR nodes, as shown in Fig.~\ref{fig:protocol}(b), generates entangled signal and idler modes that are sent to two adjacent QR nodes through two pure-loss channels, yielding a shared entangled state that is subsequently used to teleport a CV quantum state between the two QR nodes. Earlier results of CV quantum teleportation (e.g., Ref.~[86]) showed that the teleportation channel is equivalent to an additive thermal noise channel due to finite squeezing and TMSV distribution loss. The variance of additive noise is
\begin{equation}
    \sigma^2_{T}=\sqrt{\eta}
    10^{-s/10}+\left(1-\sqrt{\eta}\right),
\label{eq:p2}
\end{equation}
where $s$ (i.e. unit dB) characterizes the squeezing level of TMSV \BH{(see~\ref{sec:tele})}. 

Fig.~\ref{fig:additivenoise} plots the additive noise of the amplified one-way channel (red) and the teleportation channel (blue). Apparently, the inter-repeater spacing, $L_{\Delta}$ is a crucial factor for determining the optimal transmission protocol, and Fig.~\ref{fig:additivenoise} implies there exists a minimal inter-repeater spacing (MIRS) \begin{equation}
    L^{*}_{\Delta}\equiv-2\left(\log_{10}\left[1-10^{-s/10}\right]\right)/\gamma,
    \label{eq:criteria}
\end{equation}
such that $\sigma^2_T<\sigma^2_A$, $\forall L_{\Delta}>L^{*}_{\Delta}$.

\section{GKP-error-correction code}
\label{sec:background}
Before proceeding to GKP-assisted QEC, we will first introduce the GKP ancilla mode in Sec.~\ref{sec:GKPcodes} and the GKP-two-mode-squeezing code in Sec.~\ref{sec:GKP_TMS}.

\subsection{The GKP state}
\label{sec:GKPcodes}
A bosonic mode of, e.g., the photon or the phonon, encompasses the continuous degrees of freedom in the position and momentum quadratures. Mathematically, the quadratures, $\hat{q}$ and $\hat{p}$, are the normalized real and imaginary parts of the annihilation operator $\hat{a}$,
\begin{equation}
    \hat{q}=\frac{1}{\sqrt{2}}\left(\hat{a}+\hat{a}^{\dagger}\right),\;\;
    \hat{p}=\frac{1}{i\sqrt{2}}\left(\hat{a}-\hat{a}^{\dagger}\right),
    \label{eq:convention}
\end{equation} 
satisfying the commutation relation $\left[\hat{q},\hat{p}\right]=i$ ($\hbar\equiv1$ for simplicity). The GKP state is pure and stabilized by the following CV analog of the Pauli-Z and Pauli-X operators:
\begin{equation}
    \hat{Z}=\hat{D}\left[0,\sqrt{2\pi}\right],\;\;\hat{X}=\hat{D}\left[\sqrt{2\pi},0\right],
\end{equation}
where 
$
\hat{D}\left[\alpha,\beta\right]=e^{i\left(\alpha\hat{p}-\beta\hat{q}\right)}.
$
An ideal GKP state can be considered as the superposition of an infinite number of position or momentum eigenstates along a grid, i.e.,
\begin{equation}
     \ket{\text{GKP}}\propto\sum_{n\in\mathbb{Z}}\ket{q=n\sqrt{2\pi}}\propto\sum_{n\in\mathbb{Z}}\ket{p=n\sqrt{2\pi}}.
     \label{eq:perfectGKp}
\end{equation}

\begin{figure}[t]
	{\centering\includegraphics[width=0.6\linewidth]{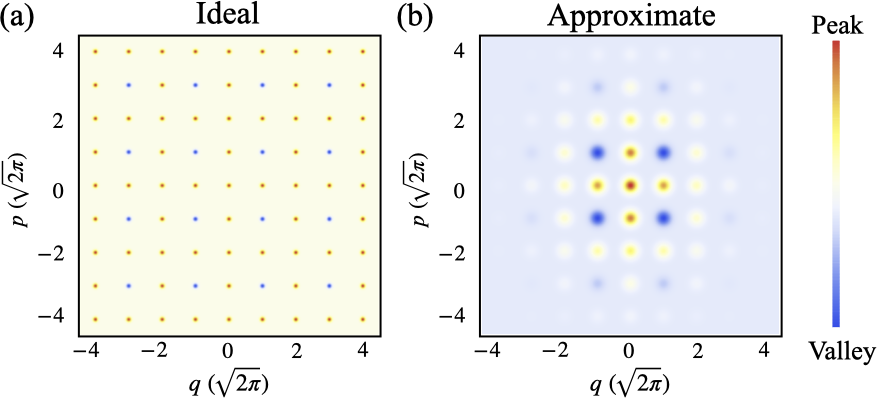}}
	\centering
	\caption{The Wigner functions in phase space of (a) ideal and (b) approximate GKP states.
	}\label{fig:G}
\end{figure}

The Wigner function of the ideal GKP state is sketched in Fig.~\ref{fig:G}(a), where each dot represents a Dirac delta function. A GKP state incorporates precise information of both quadratures within a critical range without violating the uncertainty principle. Precisely, the standard deviation of both quadrature operators modulo $\sqrt{2\pi}$ are zero. Hence, both quadratures can be measured simultaneously modulo $\sqrt{2\pi}$, rendering the GKP state perfect to calibrate any other states encoded by the GKP codes. Nonetheless, ideal GKP states are not normalizable and thus not physical. The consideration of experimental feasibility calls for a CV QEC based on approximate GKP states, as presented below.

\BH{The approximate GKP considers an uncertainty $\xi^{(\text{G})}_{q(p),2}\in\mathcal{N}\left(0,2\sigma^2_{\text{G}}\right)$ on both quadratures of each tooth. For an approximate GKP state, a series of Dirac delta functions in Eq.~\eqref{eq:perfectGKp} are replaced by a series of Gaussian packets weighted by a Gaussian profile
\begin{equation}
\begin{aligned}
    \ket{\text{GKP}}\propto&\sum_{n\in\mathbb{Z}}e^{-\pi \sigma_{\text{G}}^2n^2}\int_{-\infty}^{\infty}e^{-\frac{\left(q-\sqrt{2\pi}n\right)^2}{2\sigma_{\text{G}}^2}}\ket{q}dq\;\;\propto&\sum_{n\in\mathbb{Z}}e^{-\frac{ \sigma_{\text{G}}^2p^2}{2}}\int_{-\infty}^{\infty}e^{-\frac{\left(p-\sqrt{2\pi}n\right)^2}{2\sigma_{\text{G}}^2}}\ket{p}dp,
    \label{eq:imperfectG}
\end{aligned}
\end{equation}
and its Wigner function is plotted in Fig.~\ref{fig:G}(b)~[74,75,80,87]. The linewidths of each Gaussian teeth is characterized by the squeezing parameter $s^{(\text{G})}=-10\log_{10}\left
[2\sigma^2_{\text{G}}\right]$ (i.e. unit dB). \BH{At $\sigma_{\text{G}}\ll1$}, the Gaussian envelope can be ignored so that the approximate GKP state approaches the ideal GKP state.}

\begin{figure}[t]
	{\centering\includegraphics[width=0.45\linewidth]{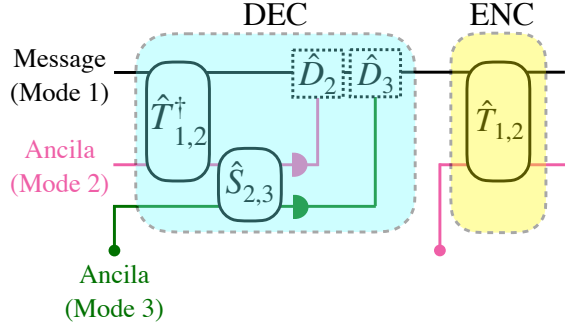}}
	\centering
	\caption{General architecture of CV QEC protocol. The light blue shaded area denotes decoding (i.e. DEC), and the light yellow shaded area denotes encoding (i.e. ENC).}
	\label{fig:encoding}
\end{figure}

\subsection{GKP-two-mode-squeezing code}
\label{sec:GKP_TMS}
The CV QEC code that is assisted with GKP state refers to GKP code and were developed to protect a bosonic mode by encoding it into multiple oscillator modes. A few of GKP codes have already been well discussed in Ref.~[67], such as, GKP-two-mode-squeezing (GKP-TMS), GKP-repetition (GKP-R) and GKP-squeezing-repetition (GKP-SR) codes, and, for consistency, the following QEC protocols all refer to the GKP-TMS code. To exploit the GKP-TMS code in the CV QR architecture, a QR node that entails an encoding operation and a decoding operation is designed, as sketched in Fig.~\ref{fig:encoding}.

To correct the additive noise, which can be modeled as independent and identically distributed (i.i.d.) Gaussian random displacements \BH{$\left(\zeta_{q,1},\zeta_{p,1},\zeta_{q,2},\zeta_{p,2}\right)$} on the four quadratures of the two modes, the \emph{encoding} process is carried out by a two-mode-squeezing (TMS) gate,  $\hat{T}_{1,2}\left[g\right]=e^{\frac{g}{2}\left(\hat{a}_{1}\hat{a}_{2}-\hat{a}^{\dagger}_{1}\hat{a}^{\dagger}_{2}\right)}$, where \BH{$g=\log\left[\sqrt{G}+\sqrt{G-1}\right]$} is determined by $G$ (i.e. $G\ge1$), and $\hat{a}_{1}$ and $\hat{a}_{2}$ denote, respectively, the annihilation operators of the involved bosonic message and ancilla mode (mode 1 and mode 2). The \emph{decoding} process entails three parts: inverse TMS operation (i.e. $\hat{T}^{\dagger}_{1,2}$), estimating the channel-induced noise by a quantum measurement and using displacement operations \BH{$\hat{D}_{2}=\hat{D}\left[-\bar{\xi}_{q,1},0\right]$ and $\hat{D}_{3}=\hat{D}\left[0,-\bar{\xi}_{p,1}\right]$} to compensate for the displacement errors incurred by the channel on the message mode, where the displacements depend on the measurement outcomes of the corresponding modes. To perform the quantum measurement, one introduces an additional GKP ancilla (mode 3). Two homodyne measurements on the prepared two ancilla modes (mode 2 and mode 3) \BH{are} implemented by a SUM gate beforehand, i.e., $\hat{S}_{2,3}=e^{-i\hat{q}_{2}\otimes\hat{p}_{3}}$). Here, $\bar{\xi}_{q,1}$ and $\bar{\xi}_{p,1}$ are the estimations of the displacement error \BH{$\xi_{q,1}=\sqrt{G}\zeta_{q,1}-\sqrt{G-1}\zeta_{q,2}$ and $\xi_{p,1}=\sqrt{G}\zeta_{p,1}+\sqrt{G-1}\zeta_{p,2}$,} acquired by measuring the ancila states in mode 2 and mode 3. \BH{In terms of experimental realization of the two in-line gates, TMS and SUM operations can be carried out via linear optics, homodyne detection, and off-line squeezers~[80,89-91].}

The corrected message mode is subsequently encoded with a new GKP ancilla at mode 2 generated at the present QR node, and both mode 1 and mode 2 are transmitted to the next QR node for decoding and encoding.

The displacement noise continuously accumulates on the message mode until it arrives at Bob's terminal. In a weak additive noise regime~[83], the displacement noise is approximately a Gaussian noise so the Wigner function of the message mode can be fully derived based on the variance of displacement noise. Let $L$ be the physical distance between Alice and Bob, the average variances of the displacement noise for Bob's received message mode are derived as
\begin{equation}
\begin{aligned}
    \Sigma^2_{QA}&=\left(L/L_{\Delta}\right)\BH{\mathcal{V}_{Q}}\left[\sigma^2_{A}\right],\;\;\Sigma^2_{QT}=\left(L/L_{\Delta}\right)\BH{\mathcal{V}_{Q}}\left[\sigma^2_{T}\right],
    \label{eq:additivenoise}
\end{aligned}
\end{equation}
over, respectively, the QEC amplified one-way and the QEC teleportation channels, where $\sigma^2_{A(T)}$ is a number given by Eq.~\eqref{eq:p1} (Eq.~\eqref{eq:p2}). Here, \BH{$\mathcal{V}_Q\left[\star\right]$} is a function to calculate the variance of the displacement noise \BH{(see \ref{sec:logicalvar})}. 

\subsection{Fidelity Performances} 
\label{sec:analysis}
This section compares the performances of CV QR with different types of quantum channels from the choices of pre- and post-processing. We will focus on the establishment of CV entanglement in the form of TMSV pairs between Alice and Bob. The overall input-output relations are constructed as the following channels: \BH{$\mathcal{T}^{D}_{L}\left[\star\right]$} for direct one-way transmission, \BH{$\mathcal{T}^{QA}_{L}\left[\star\right]$} for QEC amplified one-way transmission, and \BH{$\mathcal{T}^{QT}_{s,L}\left[\star\right]$} for QEC teleportation. In the three regimes, the GKP-TMS code is optimized over $G$ for any given parameters of the inter-repeater spacing $L_{\Delta}$, the squeezing parameter $s$ of the TMSV in quantum teleportation, and the finite squeezing teeth of the GKP state $s^{\text{(G)}}$. 

\begin{table}[]
    \centering
    \begin{tabular}{|c||c|c||c|}
    \hline
         $r$ & Message squeezing & $L$ & Spacing between Alice and Bob\\
    \hline
         $s$ & Teleportation squeezing & $L_{\Delta}$& Inter-repeater spacing\\
    \hline
    $s^{\text{(G)}}$ & GKP teeth squeezing & $L^*_{\Delta}$& MIRS s.t. $\sigma^2_T<\sigma^2_A$\\
    \hline
    \end{tabular}
    \caption{Definitions of related symbols.}
    \label{tab:summary}
\end{table}

To establish CV entanglement in the form of TMSV pairs, we focus on the following scenario: Alice generates a TMSV state consisting of a pair of modes, signal and idler, characterized by the squeezing level $r$ \BH{(in dB)}. Alice attempts to transmit the idler mode to Bob via a series of QRs while locally retaining the signal mode. In doing so, Alice and Bob share a pair of noisy TMSV. We will evaluate the performance of the QR in terms of the fidelity of the established TMSV to the ideal TMSV. The symbols of related parameters are summarized in Tab.~\ref{tab:summary}.

The Uhlmann fidelity is a measure to quantify the similarity between two density operators, $\hat{\rho}$ and $\hat{\rho}'$, defined as
\begin{equation}
    F\left[\hat{\rho},\hat{\rho}'\right]\equiv\left(\text{Tr}\left[\sqrt{\sqrt{\hat{\rho}'}\hat{\rho}\sqrt{\hat{\rho}'}}\right]\right)^2.
    \label{eq:fidelitydef}
\end{equation}
The fidelity is used to quantify the deviation between the distributed TMSV state and the original TMSV state, \BH{and can be calculated via the covariance matrices (CMs) of the involved CV quantum states (see \ref{sec:QF}).}

The fidelities of direct one-way transmission (i.e. neither pre- nor post-processing), QEC amplified one-way transmission, and QEC teleportation are defined, respectively, as
\BH{
\begin{equation}
\begin{aligned}
    F_{{O}}\equiv&F\left[\hat{\rho},\hat{\rho}'_{D}\right],\;\;F_{QA}\equiv&F\left[\hat{\rho},\hat{\rho}'_{QA}\right],\;\;F_{QT}\equiv&F\left[\hat{\rho},\hat{\rho}'_{QT}\right],
\end{aligned}
\end{equation}
where
\begin{equation}
\begin{aligned}
    \hat{\rho}'_{{D}}&=\left(\mathcal{I}\otimes \mathcal{T}^{D}_{L} \right)\left[\hat{\rho}\right],\;\;\hat{\rho}'_{QA}&=\left(\mathcal{I}\otimes \mathcal{T}^{QA}_{L} \right)\left[\hat{\rho}\right],\;\;\hat{\rho}'_{QT}&=\left(\mathcal{I}\otimes \mathcal{T}^{QT}_{s,L}\right)\left[\hat{\rho}\right].
    \label{eq:channel}
\end{aligned}
\end{equation}
} Here, $\mathcal{I}$ is the identity channel assuming ideal signal storage, and \BH{$\hat{\rho}=\ket{\text{TMSV}}\bra{\text{TMSV}}$} is the input TMSV state.

\begin{figure}
	{\centering\includegraphics[width=1\linewidth]{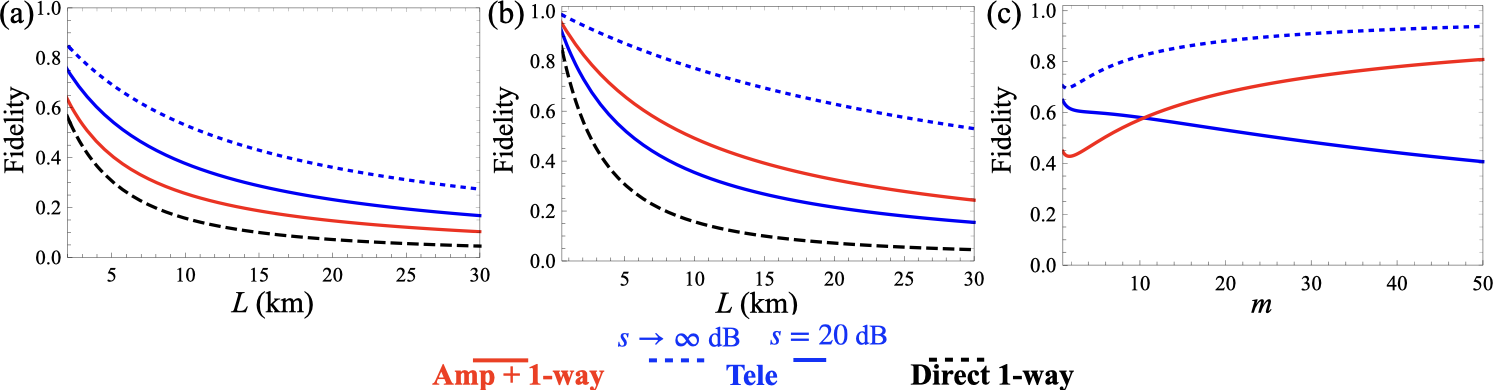}\\}
	\caption{\BH{Based on ideal GKP state}, fidelities of direct one-way transmission ($F_{O}$), QEC amplified one-way transmission ($F_{QA}$), and $m$-relay QEC teleportation ($F_{QT}$) \BH{versus $L$, with repeater spacing} (a) $L_{\Delta}=1$~km and (b) $L_{\Delta}=0.25$~km\BH{, and (c) versus numbers of repeaters $m$ at $L=5$~km}. $s^{(G)}\rightarrow\infty$, $r=15$~dB and $s=\left\{20,\infty\right\}$~dB (i.e. $L^{*}_{\Delta}=\left\{0.44,0\right\}$~km).}
	\label{fig:perfect}
\end{figure}

First, let us assume a perfect GKP state is available (i.e. $s^{(\text{G})}\rightarrow\infty$) and plot the optimized fidelities in Fig.~\ref{fig:perfect}(a) and (b). Given that the teleportation squeezing is $s=20$~dB, we choose $L_{\Delta}=250$~m to coincide with the optimal repeater separation that Rozp\c{e}dek \emph{et al.} selected in their article~[76].

The simulation result indicates that at an infinite teleportation squeezing level, i.e. $s\rightarrow\infty$, $\sigma^2_{A}>\sigma^2_{T}$ always holds, yielding $L^*_{\Delta}=0$; yet, infinite squeezing requires unbounded energy and is therefore unphysical. With a practical finite teleportation squeezing level, there is an associated non-zero MIRS. However, a shorter inter-repeater spacing increases the density of QRs and the associated resource overhead. In contrast, the QR protocol based on quantum teleportation channels reduces the density of QRs while maintaining a high fidelity for the transmitted quantum states by placing the TMSV source in the middle between two QR nodes separated by a distance of $L_{\Delta}> L^*_{\Delta}$, as shown in Fig.~\ref{fig:perfect}(b). The GKP-TMS code drastically improves the fidelity for the transmitted quantum state in both channel scenarios, as compared to the direct one-way transmission. \BH{Fig.~\ref{fig:perfect}(c) plots how the fidelity scales with the numbers of introduced repeaters $m = L/L_{\Delta} -1$.}

\begin{figure}
	{\centering\includegraphics[width=1\linewidth]{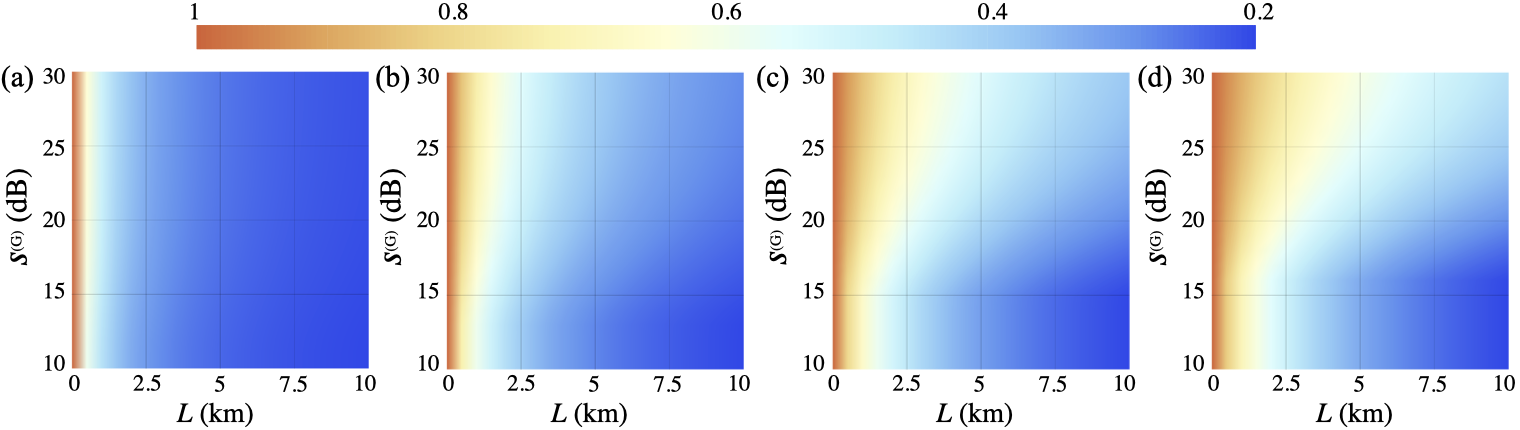}\\}
	\caption{Fidelities of \BH{QEC} teleportation-enabled repeater \BH{($F_{QT}$s)} based on imperfect GKP. Input TMSV is set $r=15$~dB and the inter-repeater separation is $L_{\Delta}=1$~km. (a) $s=10$~dB ($L^{*}_{\Delta}=4.6$~km), (b) $s=15$~dB ($L^{*}_{\Delta}=1.4$~km), (c) $s=20$~dB ($L^{*}_{\Delta}=0.44$~km), (d) $s=25$~dB ($L^{*}_{\Delta}=0.14$~km).}\label{fig:optimum}
\end{figure}

Assuming using imperfect GKP states in QEC, $F_{QT}$s are plotted in Fig.~\ref{fig:optimum}(a)(b)(c)(d) as functions of $L$ and $s^{(\text{G})}$ while fixing $r=15$~dB, corresponding to different $s$. Fig.~\ref{fig:optimum} concludes that $s^{(\text{G})}\gtrsim s\gtrsim r$ is required for effective QEC over quantum teleportation channels; otherwise, under $s<r$, the additive noise caused by teleportation will add too much noise to the transmitted quantum state while under $s^{(\text{G})}<s$, the GKP state only increases the added noise because the variance of GKP state is even larger than the noise to be corrected.

\subsection{Concatenation of GKP-TMS code}
\label{sec:concatenation}
\begin{figure}
	{\centering\includegraphics[width=0.9\linewidth]{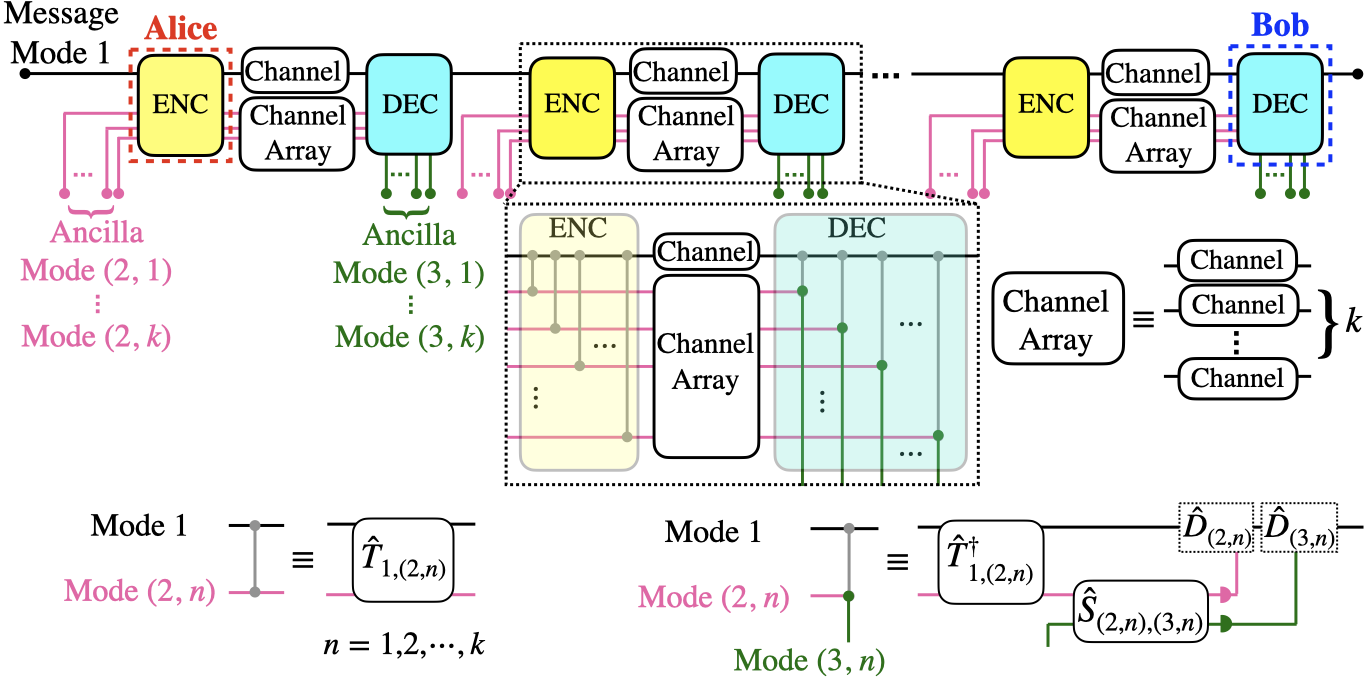}\\}
	\caption{Scheme of $m$-relay $k$-layer CV QEC repeaters. \BH{The wiring in the boxes of \emph{encoding} and \emph{decoding} are defined in the bottom. $\hat{T}_{1,\left(2,n\right)}$, is the TMS gate operated on mode 1 and mode $\left(2,n\right)$; $\hat{S}_{\left(2,n\right),\left(3,n\right)}$ is 
	the SUM gate operated on mode $\left(2,n\right)$ and mode $\left(3,n\right)$; $\hat{D}_{\left(2,n\right)}$ and $\hat{D}_{\left(3,n\right)}$ are two displacements based on the measurement outcomes of mode $\left(2,n\right)$ and mode $\left(3,n\right)$. In \emph{encoding}, mode 1 is operated by $\bigotimes^{k}_{n=1}\hat{T}_{1,\left(2,n\right)}$ with ancilla modes $\left(2,1\right)$, $\left(2,2\right)$, $\cdots,\left(2,k\right)$, and, along with these $k$ modes, distributed to the next node; in \emph{decoding}, $\bigotimes^{k}_{n=1}\hat{T}^{\dagger}_{1,\left(2,n\right)}$ is operated on the teleported $\left(k+1\right)$ modes; afterwards, $\bigotimes_{n=1}^k\hat{S}_{\left(2,n\right),\left(3,n\right)}$ is operated on the $2k$ ancilla modes, $\left(2,1\right)$, $\left(2,2\right)$, $\cdots,\left(2,k\right)$ and $\left(3,1\right)$, $\left(3,2\right)$, $\cdots,\left(3,k\right)$, for simultaneously accessing the measurement outcomes of both quadratures, and the outcomes are, ultimately, feedforwarded to mode 1.}}
	\label{fig:multi-layer}
\end{figure}
Recent study has proven that concatenation of multiple layers of QEC would substantially reduce the displacement noise comparing with only a single layer code~[83]. In a multi-layer QEC scheme, Alice, Bob and all repeaters prepare $k$ GKP ancilla (i.e. $k\in\mathbb{N}$) to be encoded with a single message state, shown in Fig.~\ref{fig:multi-layer} and \BH{another $k$ GKP ancilla to decode the teleported state. In $k$-layer QEC, the message mode in mode 1 is encoded with $k$ ancilla modes $\left(2,1\right)$, $\left(2,2\right)$, $\cdots,\left(2,k\right)$; then, the $k$-layer encoded message mode and the $k$ encoding ancilla modes are distributed to the next node over the associative channels; finally, the distributed $k+1$ modes are decoded with the another set of ancilla $\left(3,1\right)$, $\left(3,2\right)$, $\cdots,\left(3,k\right)$. As the assumption before, the physical noise of QEC can be approximately Gaussian given that the displacement noise is much less than unity~[83]. This $k$-layer QEC process corrects the aboriginal noise to the $k$-th order. In multi-layer QEC, the first layer corrects the noise with variance $\sigma^2_{0}$ carried on the received signal, yielding output noise with a variance of $\sigma^2_1=\mathcal{V}_Q\left[\sigma^2_0\right]$; the second layer then corrects the noise from the first layer QEC and results in a variance $\sigma^2_2=\mathcal{V}_Q\left[\sigma^2_1\right]$; subsequently, the $k$th-layer corrects the output noise of the $\left(k-1\right)$th-layer, leading to a residue noise variance of $\sigma_{k}^2=\mathcal{V}_Q\left[\sigma_{k-1}^2\right]$.}

\begin{figure}
	{\centering\includegraphics[width=0.45\linewidth]{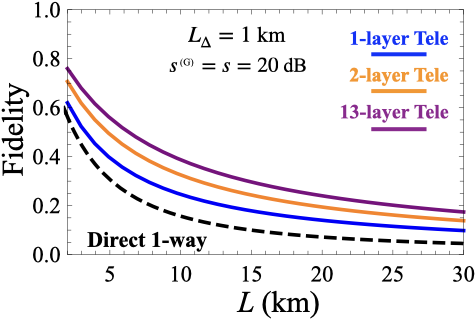}\\}
	\caption{Fidelities of $m$-relay $k$-layer ($k=1,2,13$) QEC teleportation \BH{($F_{QT}$s)}. The message squeezing is $r=15$~dB.}
	\label{fig:layer}
\end{figure}

Although the resources for implementing $m$-relay $k$-layer GKP-assisted QEC are immense (i.e. in total, $2\left(m+1\right)k$ GKP ancilla modes need to be prepared beforehand), the correction outcomes are remunerable. In Fig.~\ref{fig:layer}, we demonstrate the fidelities of the $m$-relay QEC QRs, that correspond to different layers of QEC and it shows that the fidelities are significantly improved. Albeit TMSV and GKP modes are never ideal in practice, concatenating multi-layer QEC codes is an alternative approach to suppress the additive noises of the channel, shown in Fig.~\ref{fig:layer}. In Fig.~\ref{fig:layer}, as $k\gtrsim13$, the endmost iterative noise almost converges to a finite value, which is ultimately determined by $s^{(\text{G})}$.

\section{Applications}
\label{sec:App}
Preshared entanglement between distant parties underpins numerous quantum applications. Nonetheless, establishing entanglement at a distance is impeded by the loss of the entanglement-distribution channel. The proposed GKP-assisted QEC can correct the Gaussian errors to enhanced the performance of a multitude of applications, including EA communication, QI, and CV-QKD. For simplicity, we will set $s^{(\text{G})}=s$ in the following performance analysis on the three applications assisted by the proposal QR protocol \BH{(detailed theoretical derivations are shown in~\ref{sec:formulaofapp}).}

\subsection{Entanglement-assisted communication}
\begin{figure}
	{\centering\includegraphics[width=1\linewidth]{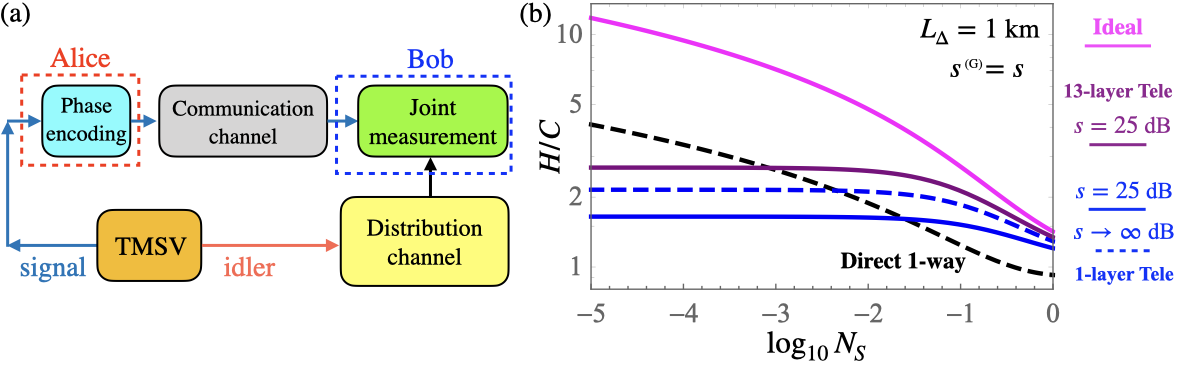}\\}
	\caption{Entanglement-assisted communication (a) scheme and (b) the phase encoding Holevo information normalized to classical capacity ($H/C$) at $L=25$~km for the cases of ideal, direct one-way transmission and $m$-relay $k$-layer ($k=1,13$) QEC teleportation.} 
	\label{fig:EAComm}
\end{figure}
The classical information rate over a thermal-loss channel is upper bounded by the classical capacity~[50,92], formulated as
\begin{equation}
    C=g\left[\kappa N_{S}+N_{B}\right]-g\left[N_{B}\right],
    \label{eq:classicalcapacity}
\end{equation}
where $g\left[x\right]\equiv\left[x+1\right]\log_{2}\left[x+1\right]-x\log_{2}x$, $N_{S}$ is the mean photon number of a signal mode, $\kappa$ is the transmissivity of the channel, and $N_{B}$ is the mean photon number of thermal-noise bath mode. EA communication is able to surpass the classical capacity~[50,93-95]. In an ideal EA communication scenario illustrated in Fig.~\ref{fig:EAComm}(a), Alice performs phase encoding on the signal mode of a preshared TMSV state and sends it to Bob over a very lossy and noisy channel, i.e., $\kappa\ll 1$ and $N_{B}\gg1$. Bob then performs a joint measurement on the received signal with the idler at hand.

However, building up preshared entanglement in real-world operational environments hinges on lossy entanglement-distribution channels that degrade the quality of the entanglement, holding back the advantage endowed by EA communication. The proposed CV QR architecture opens a promising route toward mitigating the loss arising from the entanglement-distribution channel.

The EA capacity normalized to the classical capacities are sketched as the dashed black, dashed blue, solid blue, solid purple and solid magenta curves, associated with different scenarios of entanglement sharing, in Fig.~\ref{fig:EAComm}(b). Over an extremely lossy and noisy communication channel, the asymptotic Holevo capacity normalized to the classical capacity is given by
\begin{equation}
\begin{aligned}
    H_{\text{Ideal}}/C&\approx\left(N_{S}+1\right)\log\left[1+1/N_{S}\right]\,\,,\,\,\,H_{D}/C\approx\eta\left(N_{S}+1\right)\log\left[1+1/\eta N_{S}\right],\\
    H_{QT}/C&\approx\left(N_{S}+1\right)\log\left[1+1/\Sigma^2_{QT}\right]-N_{S}/\left(\Sigma^2_{QT}+\Sigma^4_{QT}\right),
\end{aligned}
\label{eq:holevo}
\end{equation}
where $H_{\text{Ideal}}$, $H_{D}$ and $H_{QT}$ denote the Holevo information associated with ideal preshared TMSV states, TMSV sharing via direct one-way transmission and QEC teleportation-enabled QR. The QEC inevitably introduces thermal noise, causing the EA Holevo information to saturate at weak $N_{S}$'s. In this regime, teleportation is inferior to direct one-way transmission in entanglement distribution. Conversely, as $N_{S}$ increases, QEC teleportation-enabled QR starts to outperform the direct one-way entanglement distribution approach. Under this parameter setting, we find that the multi-layer encoding on finite squeezed TMSV and GKP states is more powerful than single-layer encoding on infinitely squeezed TMSV and GKP states.

\subsection{Quantum illumination}
\begin{figure}
	{\centering\includegraphics[width=0.9\linewidth]{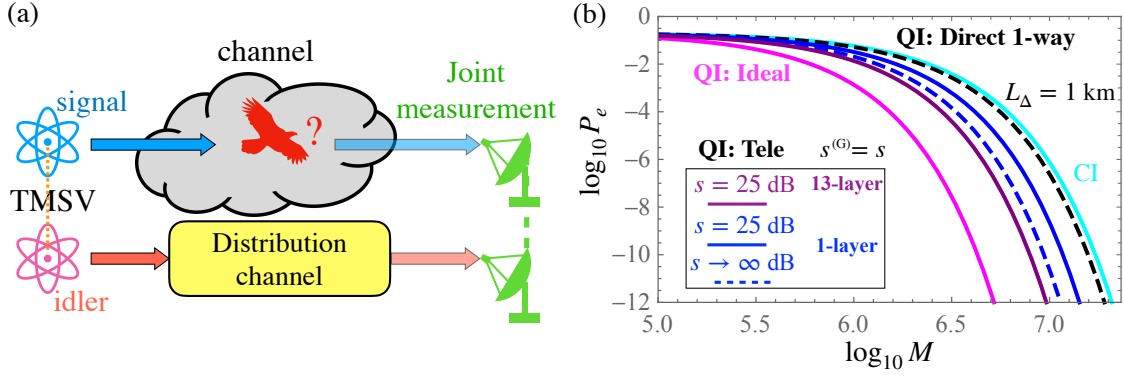}\\}
	\caption{(a) Scheme of quantum illumination. (b) The quantum Chernoff bounds of error probability $P_e$ versus transmitted $M$ modes for CI and three QI cases: ideal entanglement distribution, direct one-way, and $m$-relay $k$-layer ($k=1,13$) QEC teleportation at $L=25$~km with $N_{S}=0.01$.}
	\label{fig:QI_scheme}
\end{figure}
QI is a paradigm for quantum-enhanced target detection through a very lossy and noisy environment~[40,43,96-98]. Illustrated in Fig.~\ref{fig:QI_scheme}(a), the QI transmitter prepares TMSV states composed of entangled signal-idler mode pairs. The idler modes are distributed to receiver over a distribution channel while the signal modes are transmitted to interrogate a target residing in an environment modeled as a thermal-loss channel. The QI receiver performs a joint measurement on the transmitted signal embedded in a bright noise background and the idler to infer the presence or absence of the target. Tan {\em et al.}~[40] showed that QI, with ideal equipment and the optimum quantum receiver, achieves a 6-dB advantage in the error-probability exponent of the quantum Chernoff bound (magenta curve in Fig.~\ref{fig:QI_scheme}(b)) over that of classical illumination (CI) based on the coherent-state transmitter and homodyne receiver (cyan curve in Fig.~\ref{fig:QI_scheme}(b)).

A practical challenge for QI lies in the requirement for high-fidelity quantum memories used to match the propagation delay between the signal and idler modes. At present, QI experiments~[99] utilize low-loss optical fibers to store the idler, which mimics the one-way entanglement distribution channel. Due to the idler-storage loss, QI's advantage over CI quickly diminishes, as shown in the black dashed curve of Fig.~\ref{fig:QI_scheme}(b). The proposed QR architecture based on QEC and teleportation would constitute an effective approach to mitigate the idler-storage loss. The blue dashed and solid curves in Fig.~\ref{fig:QI_scheme}(b) depicts the simulation results for QI enhanced by QEC on the idler modes, showing reduced error probabilities as compared to QI without QEC. Akin to EA communication, in this case the multi-layer QEC with finite squeezing outperforms the single-layer QEC with infinite squeezing.

\subsection{CV quantum key distribution}
\begin{figure}
	{\centering\includegraphics[width=0.45\linewidth]{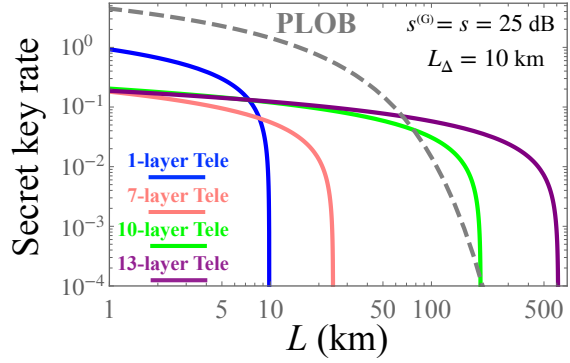}\\}
	\caption{The secret key rates per mode of $m$-relay $k$-layer ($k=1,7,10,13$) QEC teleportation-enabled QR.}
	\label{fig:Security}
\end{figure}

CV-QKD enables two distant parties, Alice and Bob, to securely share a common binary random key despite the adversary, Eve, mounts the optimal attack to capture the communicated information~[6,100-102]. Unlike its DV counterpart, CV-QKD can be fully implemented with off-the-shelf telecommunication components without resorting to single-photon detectors and is thus particularly intriguing for real-world deployment. The security of CV-QKD protocols is analyzed by upper bounding the accessible information to Eve assuming her power is only limited by the laws of physics. Specifically, the secret-key rate (SKR) for CV-QKD is given by 
\begin{equation}
    R\approx-\frac{1}{2}\log_2{\left[e^2\epsilon\left(1+\epsilon\right)/4\right]},
    \label{eq:securitykeyrate}
\end{equation}
where $\epsilon$ quantifies the variance of overall additive excess noise. The proposed QR architecture based on $m$-relay $k$-layer QEC mitigates the loss of the quantum channel to boost the SKR, as shown in Fig.~\ref{fig:Security}.

To further investigate the application of the QR architecture to CV-QKD, two additional remarks on Fig.~\ref{fig:Security} are worth making. First, the SKR of the QR architecture based on $k$-layer QEC and teleportation are below the PLOB bound at $k \leq 9$, hindered by the accumulated noise introduced at the QR nodes. Second, given $s=s^{(\text{G})}=25$~dB, the theoretical maximal distance of the QR architecture based on 13-layer QEC and teleportation, as shown in the purple curve of Fig.~\ref{fig:Security}, reaches 596~km. We expect that the incorporation of an additional DV QEC layer would suppress the residue noise and further extend the CV-QKD distance~[76,77].

\section{Discussion and outlook}\label{sec:Diss}
The QR architecture based on teleportation channels places an entanglement source in the middle of two adjacent QR nodes. In contrast, the QR scheme based on amplified one-way channels directly connects the adjacent nodes by optical fibers. One may argue that adding an intermediate QR node in \BH{an} amplified one-way channel would surpass the performance of the teleportation-based scheme. However, a full-scale QR node needs multiple GKP ancilla modes, which consumes much more resources than the widely available TMSV source.

The combination of CV and DV QEC was recently proposed by Rozp\c{e}dek \emph{et al.}~[76] and Fukui \emph{et al.}~[77]. Such a hybrid QEC scheme would allow the proposed QR architecture based on $m$-relay $k$-layer QEC to be further concatenated with a DV QEC code to drastically reduce the amount of residue noise. As long as the CV errors after QEC are limited to a certain range, the DV QEC will be able to correct these errors to maximize the fidelity of \BH{the transmitted quantum state}.
\\

\section{Conclusions}\label{sec:conc}
In this article, we proposed a deterministic CV QR architecture based on \BH{optical} GKP states to enable the distribution of CV quantum states over long distances. \BH{The proposed QR architecture based on GKP QEC obviates the needs for quantum memories and thereby remarkably reduces the burden on quantum information storage; moreover, it significantly suppresses the additive errors caused by a lossy bosonic channel}. In our study, we showed that the \BH{optical} QR architecture based on GKR QEC and teleportation outperforms direct one-way transmission when the squeezing level is higher than 15 dB. The proposed QR architecture is applied to improve the performance of EA communication, QI and CV-QKD. Once \BH{optical} GKP states with sufficient squeezing become available, the proposed QR architecture will enable CV quantum states to be faithfully transmitted over unprecedented distances, thereby making a large stride forward in the development of quantum technology.

\ack
This research is supported by National Science Foundation Grant No. ECCS-1920742, CCF-1907918, and No. EEC-1941583, and Defense Advanced Research Projects Agency (DARPA) under Young Faculty Award (YFA) Grant No. N660012014029.

\appendix

\section{Additive Gaussian noises of protocols} 
In this section, we prove that both teleportation and QEC result in additive zero-mean Gaussian noises to the quantum system. 
\subsection{Teleportation}
\label{sec:tele}
\begin{figure}
	{\centering\includegraphics[width=0.9\linewidth]{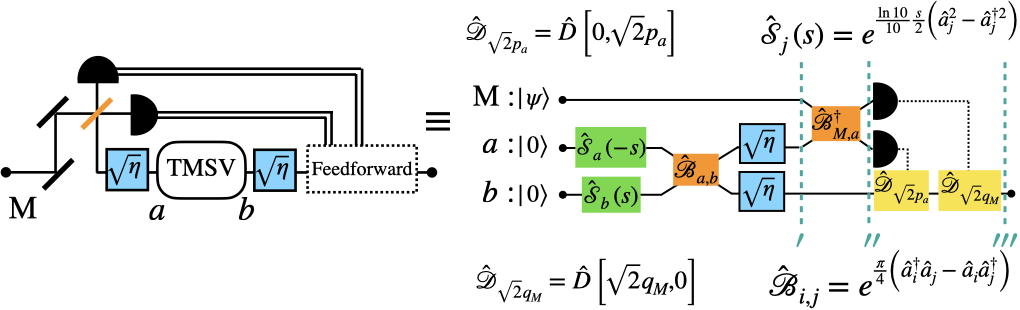}\\}
	\caption{The quantum circuit scheme of teleportation. The apostrophes, $'$, $''$ and $'''$, stand for the stages, mentioned in the context. $q_M=\langle\hat{q}^{\text{(T)}''}_{M}\rangle$ and $p_a=\langle\hat{p}^{\text{(T)}''}_{a}\rangle$.
	\label{fig:telescheme}
	}
\end{figure}
The quantum circuit of teleportation is shown in Fig.~\ref{fig:telescheme}. In teleportation, sender prepares multiple TMSV states (with quadratures $\hat{q}^{\text{(T)}}$ and $\hat{p}^{\text{(T)}}$) at the middle of two consecutive nodes. The off-line TMSV state have the quadratures
\begin{equation}
    \begin{aligned}
        \hat{q}^{(\text{T})}_{a}&=\left(\hat{q}_{a}^{(\text{v})}10^{s/20}+\hat{q}_{b}^{(\text{v})}10^{-s/20}\right)/\sqrt{2},\;\;
        \hat{p}^{(\text{T})}_{a}=\left(\hat{p}_{a}^{(\text{v})}10^{-s/20}+\hat{p}_{b}^{(\text{v})}10^{s/20}\right)/\sqrt{2},\\
        \hat{q}^{(\text{T})}_{b}&=\left(\hat{q}_{a}^{(\text{v})}10^{s/20}-\hat{q}_{b}^{(\text{v})}10^{-s/20}\right)/\sqrt{2},\;\;
        \hat{p}^{(\text{T})}_{b}=\left(\hat{p}_{a}^{(\text{v})}10^{-s/20}-\hat{p}_{b}^{(\text{v})}10^{s/20}\right)/\sqrt{2}
    \end{aligned}
    \label{eq:TMSVq}
\end{equation}
for submodes $a$ and $b$, where $\hat{q}^{(\text{v})}$ denotes the vacuum operator. In Eq.~\ref{eq:TMSVq}. The submodes $a$ and $b$ are distributed, respectively, to the former nodes and the later one. Since TMSV is put in the middle, the distribution channel becomes two sub-channels with transmissivity $\eta^{1/2}$ and the attenuated quadratures become,
\begin{equation}
    \begin{aligned}
        \hat{q}_{a}^{(\text{T})'}&=\sqrt{\eta^{1/2}}\hat{q}^{(\text{T})}_{a}+\sqrt{1-\eta^{1/2}}\hat{q}_{a}^{(\text{v})'},\;\;
        \hat{p}_{a}^{(\text{T})'}=\sqrt{\eta^{1/2}}\hat{p}^{(\text{T})}_{a}+\sqrt{1-\eta^{1/2}}\hat{p}_{a}^{(\text{v})'},\\
        \hat{q}^{(\text{T})'}_{b}&=\sqrt{\eta^{1/2}}\hat{q}^{(\text{T})}_{b}+\sqrt{1-\eta^{1/2}}\hat{q}_{b}^{(\text{v})'},\;\;
        \hat{p}^{(\text{T})'}_{b}=\sqrt{\eta^{1/2}}\hat{p}^{(\text{T})}_{b}+\sqrt{1-\eta^{1/2}}\hat{p}_{b}^{(\text{v})'},
    \end{aligned}
    \label{eq:subchannel}
\end{equation}
where $\hat{p}_{a(b)}^{(\text{v})'}$ is the transmission-induced vacuum operator at $a$ ($b$). In teleportation, sender implements the Bell measurement on $M$ (with quadratures $\hat{q}_{\text{M}}$, $\hat{p}_{\text{M}}$) and $a$, and results in the quadratures as
\begin{equation}
    \begin{aligned}
        \hat{q}^{(\text{T})''}_{a}&=\left(\hat{q}_{M}+\hat{q}^{(\text{T})'}_{a}\right)/\sqrt{2},\;\;\hat{p}^{(\text{T})''}_{a}=\left(\hat{p}_{M}+\hat{p}^{(\text{T})'}_{a}\right)/\sqrt{2},\\
        \hat{q}^{(\text{T})''}_{M}&=\left(\hat{q}_{M}-\hat{q}^{(\text{T})'}_{a}\right)/\sqrt{2},\;\;\hat{p}^{(\text{T})''}_{M}=\left(\hat{p}_{M}-\hat{p}^{(\text{T})'}_{a}\right)/\sqrt{2}.
    \end{aligned}
    \label{eq:finalstep}
\end{equation}
Subsequently, the sender feedforward the measurement results in mode $M$ and $a$ to $b$. With Eq.~\ref{eq:TMSVq}, Eq.~\ref{eq:subchannel} and Eq.~\ref{eq:finalstep}, the resulting quadratures in $b$ are
\begin{equation}
    \begin{aligned}
        \hat{q}^{(\text{T})'''}_{b}&=\hat{q}_{\text{M}}-\sqrt{2\eta^{1/2}}10^{-s/20}\hat{q}_{b}^{(\text{v})}+\sqrt{1-\eta^{1/2}}\left(\hat{q}_{b}^{(\text{v})'}-\hat{q}_{a}^{(\text{v})'}\right),\\
        \hat{p}^{(\text{T})'''}_{b}&=\hat{p}_{\text{M}}+\sqrt{2\eta^{1/2}}10^{-s/20}\hat{p}_{a}^{(\text{v})}+\sqrt{1-\eta^{1/2}}\left(\hat{p}_{b}^{(\text{v})'}+\hat{p}_{a}^{(\text{v})'}\right),
    \end{aligned}
    \label{eq:awgn}
\end{equation}
and, apparently, we acquire the formula of additive noise as in Eq.~\ref{eq:p2}.

\subsection{QEC protocol}
The QEC protocol consists of two parts: \emph{encoding} and \emph{decoding}.
\label{sec:logicalvar}
\subsubsection{Encoding}
In GKP-TMS code, we implement $\hat{T}_{1,2}\left[g\right]$ to correlate the message mode (with density operator $\hat{\rho}_{1}$) in mode 1 and an approximate GKP ancilla mode (with density operator $\hat{\rho}^{(\text{G})}_{2}$) in mode 2 as
\begin{equation}
    \hat{T}^{\dagger}_{1,2}\left[g\right]\,\left[\hat{\rho}_{1}\otimes\hat{\rho}^{(\text{G})}_{2}\right]\,\hat{T}_{1,2}\left[g\right].
\end{equation}
This TMS gate is described by a symplectic matrix,
\begin{equation}
\bold{S}_{1,2}=\begin{pmatrix}
        \sqrt{G}\,\bold{I}_{2}&\sqrt{G-1}\,\bold{Z}_{2}\\
        \sqrt{G-1}\,\bold{Z}_{2}&\sqrt{G}\,\bold{I}_{2}\\
    \end{pmatrix}
\end{equation}
in the basis of $\left(q_{1},p_{1},q_{2},p_{2}\right)^{T}$, where $\bold{I}_{n}$ and $\bold{Z}_{n}$ denote the $n\times n$ identity and Pauli Z matrices. 

\subsubsection{Decoding}
During quantum information processing (e.g. teleportation), the quantum state is added with Gaussian noises $\left(\zeta_{q,1},\zeta_{p,1},\zeta_{q,2},\zeta_{p,2}\right)\sim_{\text{iid}} \mathcal{N}\left(0,\sigma^2 \right)$ ($\sigma\in\mathbb{R}$). The noise can be characterized by the CM $\bold{V}=\sigma^2\bold{I}_{4}$. The CM, after being operated by $\bold{S}_{1,2}^{-1}$, becomes
\begin{equation}
\begin{aligned}
    \bold{S}_{1,2}^{-1}\bold{V}\left(\bold{S}_{1,2}^{-1}\right)^{\text{T}}=\begin{pmatrix}
        \left(2G-1\right)\,\bold{I}_{2}&-2\sqrt{G(G-1)}\,\bold{Z}_{2}\\
        -2\sqrt{G(G-1)}\,\bold{Z}_{2}&\left(2G-1\right)\,\bold{I}_{2}\\
    \end{pmatrix}\sigma^2,
\end{aligned}\label{eq:CM}
\end{equation}
and the formulated additive noises in mode 1 and mode 2 are:
\begin{equation}
    \begin{pmatrix}
        \xi_{q,1}\\
        \xi_{p,1}
    \end{pmatrix}
        =
    \begin{pmatrix}
    \sqrt{G}\zeta_{q,1}-\sqrt{G-1}\zeta_{q,2}\\
    \sqrt{G}\zeta_{p,1}+\sqrt{G-1}\zeta_{p,2}.
    \end{pmatrix},\;\;
    \begin{pmatrix}
        \xi_{q,2}\\
        \xi_{p,2}
    \end{pmatrix}
    =
    \begin{pmatrix}
    \sqrt{G}\zeta_{q,2}-\sqrt{G-1}\zeta_{q,1}\\
    \sqrt{G}\zeta_{p,2}+\sqrt{G-1}\zeta_{p,1}
    \end{pmatrix}.
\label{eq:correlatedrandom}
\end{equation}

At this stage, the noise $\xi_{q(p),1}$ is correlated with $\xi_{q(p),2}$ (see Eq.~\ref{eq:CM}), and, hence, can be inferred by measuring the ancilla in mode 2. Based on the minimum variance unbiased estimation (MVUE) (with a Gaussian approximation)~[67], the estimator of $\xi_{q(p),1}$, is formulated as
\begin{equation}
\begin{aligned}
    \bar{\xi}_{q,1}&=\;\text{argmin}_{\tilde{\xi}_{q,1}\in\mathbb{R}}\left\{\text{Var}\left[\xi_{q,1}-\tilde{\xi}_{q,1}\right]\right\}=-\frac{2\sqrt{G\left(G-1\right)}\sigma^2}{\left(2G-1\right)\sigma^2+2\sigma_{\text{G}}^2}R_{\sqrt{2\pi}}\left[\xi_{q,2}+\xi^{(\text{G})}_{q,2}\right],\\
    \bar{\xi}_{p,1}&=\;\text{argmin}_{\tilde{\xi}_{p,1}\in\mathbb{R}}\left\{\text{Var}\left[\xi_{p,1}-\tilde{\xi}_{p,1}\right]\right\}=\frac{2\sqrt{G\left(G-1\right)}\sigma^2}{\left(2G-1\right)\sigma^2+2\sigma_{\text{G}}^2}R_{\sqrt{2\pi}}\left[\xi_{p,2}+\xi^{(\text{G})}_{p,2}\right],
    \label{eq:corrected}
\end{aligned}
\end{equation}
where, $\text{V}\left[\star\right]$ denotes variance, $R_{\sqrt{2\pi}}\left[x\right]=x-\sqrt{2\pi}\times\text{argmin}_{n\in\mathbb{Z}}\big{|}x-\sqrt{2\pi}n\big{|}$. The state in mode 1, then, is implemented with two displacement operations $\hat{D}\left[-\bar{\xi}_{q,1},0\right]$ and $\hat{D}\left[0,-\bar{\xi}_{p,1}\right]$ to have the resulting noise
\begin{equation}
\begin{aligned}
    \xi_{q,1}-\bar{\xi}_{q,1}&=\xi_{q,1}+\frac{2\sqrt{G\left(G-1\right)}\sigma^2}{\left(2G-1\right)\sigma^2+2\sigma_{\text{G}}^2}R_{\sqrt{2\pi}}\left[\xi_{q,2}+\xi_{q,2}^{\text{(G)}}\right],\\ \xi_{p,1}-\bar{\xi}_{p,1}&=\xi_{p,1}-\frac{2\sqrt{G\left(G-1\right)}\sigma^2}{\left(2G-1\right)\sigma^2+2\sigma_{\text{G}}^2}R_{\sqrt{2\pi}}\left[\xi_{p,2}+\xi_{p,2}^{\text{(G)}}\right].
\end{aligned}
\end{equation}
When the noise is small, we can approximate $R_{\sqrt{2\pi}}\left[\xi_{q(p),2}+\xi_{q(p),2}^{\text{(G)}}\right]$ as a Gaussian random variable and therefore our QEC protocol approximately produces a Gaussian state, and we show the derivation of resulting variance after QEC in the following section.

\subsubsection{Full derivation of $\Sigma_Q^2$} 
With Eq.~\ref{eq:correlatedrandom}, the resulting variance of both quadratures are
\begin{equation}
    \begin{aligned}
        \langle\hat{q}^2\rangle=\sum^{\infty}_{n=-\infty}&\int_{-\infty}^{\infty}d\xi^{(\text{G})}_{q,2}\int_{-\infty}^{\infty}d\xi_{q,2}\int_{-\infty}^{\infty}d\xi_{q,1}\;\left[\frac{1}{\sqrt{4\pi}\sigma_{G}}e^{-\frac{\xi_{q,2}^{(\text{G})2}}{4\sigma^2_{G}}}\right]\left[\frac{1}{2\pi\sigma^2}e^{-\frac{2G-1}{2\sigma^2}\left(\xi^{2}_{q,1}+\xi^{2}_{q,2}\right)-\frac{2\sqrt{G(G-1)}\xi_{q,1}\xi_{q,2}}{\sigma^2}}\right]\times\\
        &\left(\xi_{q,1}-\bar{\xi}_{q,1}\right)^2\mathcal{U}\left(\xi_{q,2}+\xi^{(\text{G})}_{q,2}\in\left[\left(n-\frac{1}{2}\right)\sqrt{2\pi},\left(n+\frac{1}{2}\right)\sqrt{2\pi}\right]\right),\\
        \langle\hat{p}^2\rangle=\sum^{\infty}_{n=-\infty}&\int_{-\infty}^{\infty}d\xi^{(\text{G})}_{p,2}\int_{-\infty}^{\infty}d\xi_{p,2}\int_{-\infty}^{\infty}d\xi_{p,1}\;\left[\frac{1}{\sqrt{4\pi}\sigma_{G}}e^{-\frac{\xi_{p,2}^{(\text{G})2}}{4\sigma^2_{G}}}\right]\left[\frac{1}{2\pi\sigma^2}e^{-\frac{2G-1}{2\sigma^2}\left(\xi^{2}_{p,1}+\xi^{2}_{p,2}\right)+\frac{2\sqrt{G(G-1)}\xi_{p,1}\xi_{p,2}}{\sigma^2}}\right]\times\\
        &\left(\xi_{p,1}-\bar{\xi}_{p,1}\right)^2\mathcal{U}\left(\xi_{p,2}+\xi^{(\text{G})}_{p,2}\in\left[\left(n-\frac{1}{2}\right)\sqrt{2\pi},\left(n+\frac{1}{2}\right)\sqrt{2\pi}\right]\right),
    \end{aligned}
    \label{eq:logicalnoiseint}
\end{equation}
where $\mathcal{U}$ is an indicator function (i.e. $\mathcal{U}\left(\mathcal{S}\right)=1$, if $\mathcal{S}$ is true; otherwise, $\mathcal{U}\left(\mathcal{S}\right)=0$). Performing partial integration, we obtain
\begin{equation}
    \begin{aligned}
        \langle\hat{q}^2\rangle&=\sum^{\infty}_{n=-\infty}\int_{-\infty}^{\infty}d\xi^{(\text{G})}_{q,2}\int_{-\infty}^{\infty}d\xi_{q,2}\;\;e^{-\frac{\xi^2_{q,2}}{2\left(2G-1\right)\sigma^2}-\frac{\xi_{q,2}^{(\text{G})2}}{4\sigma^2_{G}}}\;\mathcal{U}\left(\xi_{q,2}+\xi^{(\text{G})}_{q,2}\in\left[\left(n-\frac{1}{2}\right)\sqrt{2\pi},\left(n+\frac{1}{2}\right)\sqrt{2\pi}\right]\right)\\
        &\times\left\{\frac{\sigma }{\left[2\left(2G-1\right)\right]^{3/2}\pi\sigma_{G}}+\frac{\sqrt{2}G\left(G-1\right)\left[\left(2G-1\right)\left(n\sqrt{2\pi}-\xi^{(\text{G})}_{q,2}\right)\sigma^2+2\xi_{q,2}\sigma^2_{G}\right]^2}{\left(2G-1\right)^{5/2}\pi\sigma_{G}\left[\left(2G-1\right)\sigma^2+2\sigma^2_{G}\right]^2\sigma}\right\}\\
        &=\sum^{\infty}_{n=-\infty}\left\{\frac{\sigma^2\left[8\left(G-1\right)Gn^2\pi\sigma^2+\left(2G-1\right)\sigma^4+4\left(2G\left(G-1\right)+1\right)\sigma^2\sigma^2_{G}+4\left(2G-1\right)\sigma^4_{G}\right]}{2\left[\left(2G-1\right)\sigma^2+2\sigma^2_{G}\right]^{2}}\right\}\\
        &\times\left\{\text{erfc}\left[\frac{\left(n-1/2\right)\sqrt{\pi}}{\sqrt{\left(2G-1\right)\sigma^2+2\sigma^2_{G}}}\right]-\text{erfc}\left[\frac{\left(n+1/2\right)\sqrt{\pi}}{\sqrt{\left(2G-1\right)\sigma^2+2\sigma^2_{G}}}\right]\right\}\equiv\mathcal{V}_{Q}\left[\sigma^2\right]=\langle\hat{p}^2\rangle.
    \end{aligned}
\end{equation}

\section{Quantum fidelities of TMSV}
\label{sec:QF}
Starting with Eq.~\ref{eq:fidelitydef}, the fidelity between two-mode Gaussian quantum states $\hat{\rho}$ and $\hat{\rho}'$ can be obtained as~[96],
\begin{equation}
    F=\left(\sqrt{\Gamma}+\sqrt{\Omega}-\sqrt{\left(\sqrt{\Gamma}+\sqrt{\Omega}\right)^2-\Theta}\right)^{-1}\exp{\left[-\frac{1}{2}\delta \bold{v}^{\text{T}}\left(\bold{C}+\bold{C'}\right)^{-1}\delta \bold{v}\right]},
    \label{eq:fidelity}
\end{equation}
where 
\begin{equation}
\begin{aligned}
    \bold{J}=\bigoplus^2_{n=1}\begin{pmatrix}0&1\\-1&0
    \end{pmatrix},\;\;\Gamma&=2^4\text{Det}\left[\bold{J}\bold{C}\bold{J}\bold{C}'-\frac{1}{4}\bold{I}_{4}\right],\;\;\;\;
    \Theta=\text{Det}\left[\bold{C}+\bold{C}'\right],\\
    \Omega&=2^4\text{Det}\left[\bold{C}+\frac{i}{2}\bold{J}\right]\text{Det}\left[\bold{C'}+\frac{i}{2}\bold{J}\right].
    \label{eq:ultimatefidelity}
\end{aligned}
\end{equation}
Here $\delta\bold{v}=\langle\bold{v}\rangle-\langle\bold{v}'\rangle$, with $\langle\bold{v}\rangle$ and $\langle\bold{v}'\rangle$ being the quadrature means of quantum states $\hat{\rho}$ and $\hat{\rho}'$, with the associated CMs $\bold{C}$ and $\bold{C}'$. Since our concerning state is zero-mean Gaussian (i.e. $\delta\bold{v}=0$), we can derive quantum fidelity by its CM. Defining $r_e=\left(r\log{10}\right)/10$, we have the CM of ideal TMSV,
\begin{equation}
    \bold{C}=\frac{1}{2}\begin{pmatrix}
    \cosh{r_e}\;\bold{I}_{2}&\sinh{r_e}\;\bold{Z}_{2}\\
    \sinh{r_e}\;\bold{Z}_{2}&\cosh{r_e}\;\bold{I}_{2}\end{pmatrix},
    \label{eq:idealtmsvcm}
\end{equation}
and CMs of the distribution channels of direct one-way, QEC amplified one-way and $m$-relay QEC teleportation,
\begin{equation}
\begin{aligned}
    \bold{C}'_{\text{D}}&=\frac{1}{2}\begin{pmatrix}
    \cosh{r_e}\;\bold{I}_{2}&\sqrt{\eta}\sinh{r_e}\;\bold{Z}_{2},\\
    \sqrt{\eta}\sinh{r_e}\;\bold{Z}_{2}&\left\{\eta\cosh{r_e}+1-\eta\right\}\;\bold{I}_{2}
    \end{pmatrix},\;\;    \bold{C}'_{\text{QA}}=\frac{1}{2}\begin{pmatrix}
    \cosh{r_e}\;\bold{I}_{2}&\sinh{r_e}\;\bold{Z}_{2}\\
    \sinh{r_e}\;\bold{Z}_{2}&\left\{\cosh{r_e}+2\Sigma^2_{QA}\right\}\;\bold{I}_{2}
    \end{pmatrix},\\
    \bold{C}'_{\text{QT}}&=\frac{1}{2}\begin{pmatrix}
    \cosh{r_e}\;\bold{I}_{2}&\sinh{r_e}\;\bold{Z}_{2}\\
    \sinh{r_e}\;\bold{Z}_{2}&\left\{\cosh{r_e}+2\Sigma^2_{QT}\right\}\;\bold{I}_{2}
    \end{pmatrix},
\end{aligned}
    \label{eq:threeCM}
\end{equation}
and use Eq.~\ref{eq:fidelity} to derive the fidelities,
\begin{equation}
    \begin{aligned}
        F_D&=\frac{4}{\left[\left(1+\sqrt{\eta}\right)+\left(1-\sqrt{\eta}\right)\cosh{r_e}\right]^2},\;\;
        F_{QA}=\frac{1}{1+\Sigma^2_{QA}\cosh{r_e}},\;\;
        F_{QT}&=\frac{1}{1+\Sigma^2_{QT}\cosh{r_e}}.
    \end{aligned}
\end{equation}

\section{Theoretical formula of applications}
\label{sec:formulaofapp}
In this section, we discuss the outcomes of three applications, considering a lossy and noisy idler distribution channel (i.e. direct one-way transmission channel), and their boosted performances after QEC process. To be consistent with the widely used quadrature convention of these applications, we choose the quadrature convention: $\hat{q}=\hat{a}+\hat{a}^{\dagger}$ and $\hat{p}=\left(\hat{a}-\hat{a}^{\dagger}\right)/i$ in the following calculations.

\subsection{Entanglement assisted communications}
In EA communication scenario, the signal arm of the prepared TMSV quantum state (with density operator $\hat{\rho}$) is encoded by a phase modulation operator $\hat{U}_{\theta}=\exp\left[i\theta\hat{a}_S^{\dagger}\hat{a}_S\right]$ (i.e. $\theta\in\left[0,2\pi\right)$) for message encoding (i.e. $\hat{\rho}_{\theta}=\hat{U}_\theta\hat{\rho}\hat{U}^{\dagger}_\theta$) to obtain the CM,
\begin{equation}
    \Lambda=\begin{pmatrix}
        \left(2N_S+1\right)\bold{I}_2&2C_0\bold{R}_\theta\\
        2C_0\bold{R}_\theta&\left(2N_S+1\right)\bold{I}_2
    \end{pmatrix},
    \label{eq:idealCM}
\end{equation}
where $C_0=\sqrt{N_S\left(N_S+1\right)}$, and $\bold{R}_\theta=\text{Re}\left\{\exp{\left[i\theta\left(\bold{Z}_2-i\bold{X}_2\right)\right]}\right\}$, $N_S$ is the mean photon number of the preshared TMSV. Here $\bold{I}_2$, $\bold{Z}_2$ and $\bold{X}_2$ are the Pauli matrices. After encoding, the signal mode is distributed to receiver via a lossy and noisy channel (i.e transmissivity $\kappa \ll1$ and $N_B=\langle\hat{a}^{\dagger}_{B}\hat{a}_{B}\rangle/\left(1-\kappa\right)$ is the mean photon number of heat bath). Given that the optimal decoding approach is applied, the Holevo (maximally accessible) capacity can be obtained from the formula
\begin{equation}
    \chi=S\left[\frac{1}{2\pi}\int^{2\pi}_{0}\hat{\rho}_\theta d\theta\right]-\frac{1}{2\pi}\int^{2\pi}_{0}S\left[\hat{\rho}_\theta\right]d\theta,
    \label{eq:holevo}
\end{equation}
where $S\left[\star\right]$ is the von Neumann entropy. As the signal and idler mode of the phase-encoded field are, respectively, transmitted to the lossy and noisy channel and the distribution channel, the CMs become
\begin{equation}
\begin{aligned}
    \Lambda_{\text{ideal}}&=\begin{pmatrix}
        \left(2N_B+2\kappa N_S+1\right)\bold{I}_2&2\sqrt{\kappa}C_0\bold{R}_\theta\\
        2\sqrt{\kappa}C_0\bold{R}_\theta&\left(2N_S+1\right)\bold{I}_2
    \end{pmatrix},\;\;
    \Lambda_D=\begin{pmatrix}
        \left(2N_B+2\kappa N_S+1\right)\bold{I}_2&2\sqrt{\eta\kappa}C_0\bold{R}_\theta\\
        2\sqrt{\eta\kappa}C_0\bold{R}_\theta&\left(2\eta N_S+1\right)\bold{I}_2
    \end{pmatrix},\\
    \Lambda_{QT}&=\begin{pmatrix}
        \left(2N_B+2\kappa N_S+1\right)\bold{I}_2&2\sqrt{\kappa}C_0\bold{R}_\theta\\
        2\sqrt{\kappa}C_0\bold{R}_\theta&\left(2N_S+2\Sigma^2_{QT}+1\right)\bold{I}_2
    \end{pmatrix},
    \label{eq:CMsEAcomm}
\end{aligned}
\end{equation}
with respect to ideal ($\Lambda_{\text{ideal}}$), direct-one way ($\Lambda_D$) and $m$-relay QEC teleportation ($\Lambda_{QT}$) distribution channel. Finally, Eq.~\ref{eq:holevo} and Eq.~\ref{eq:CMsEAcomm} allow us to calculate the Holevo capacities (more detailed calculations can be found in~[50]),
\begin{equation}
    \begin{aligned}
    \chi_{\text{ideal}}&\approx\frac{\kappa}{N_B} N_S\left(N_S+1\right)\log_2{\left[1+1/N_S\right]},\;\;\;\;\;\;\;\;
    \chi_D\approx\frac{\eta\kappa}{N_B} N_S\left(N_S+1\right)\log_2{\left[1+1/\eta N_S\right]},\\
    \chi_{QT}&\approx\frac{\kappa N_S\left\{\left(N_S+1\right)\Sigma^2_{QT}\left(\Sigma^2_{QT}+1\right)\log{\left[1+1/\Sigma^2_{QT}\right]}-N_S\right\}}{N_B\Sigma^2_{QT}\left(\Sigma^2_{QT}+1\right)\log{2}},
    \end{aligned}
    \label{eq:holevoequations}
\end{equation}
by assuming $N_S\ll1$, $\kappa\ll1$ and $N_{B}\gg1$.

\subsection{Quantum illumination}
The error probability of binary hypothesis testing in a quantum system can be evaluated from the two density operators involved in the hypotheses,
\begin{equation}
\begin{aligned}
    \text{\emph{Hypothesis 1:}}&\;\;\hat{\rho}_{1},\;\;\text{when target is present,}\\
    \text{\emph{Hypothesis 2:}}&\;\;\hat{\rho}_{2},\;\;\text{when target is absent.}
\end{aligned}
\end{equation}
With multiple copies of the unknown state, the error probability is upper bounded by the Quantum Chernoff bound (QCB)~[96],
\begin{equation}
    \frac{1}{2}\left(\inf_{0\le \nu\le1}\left\{\text{Tr}\left[\hat{\rho}_1^{\nu}\hat{\rho}_2^{1-\nu}\right]\right\}\right)^M,
    \label{eq:QCB}
\end{equation}
where $M$ is the number of identical copy of the quantum system. Ref.~[96] guides us the formula of QCB,
\begin{equation}
    \frac{1}{2}\inf_{0\le \nu\le1}\left\{\frac{2^n\prod_{j=1}^n G_\nu\left[\lambda_{1,j}\right]G_{1-\nu}\left[\lambda_{2,j}\right]}{\sqrt{\det\left[\bold{V}_1\left[\nu\right]+\bold{V}_2\left[1-\nu\right]\right]}}\exp{\left[-\frac{1}{2}\delta \bold{v}^{\text{T}}\left(\bold{V}_1\left[\nu\right]+\bold{V}_2\left[1-\nu\right]\right)^{-1}\delta \bold{v}\right]}\right\}^M,
\label{eq:QCBformula}
\end{equation}
where $G_\nu\left[x\right]=2^\nu/\left[\left(x+1\right)^\nu-\left(x-1\right)^\nu\right]$, $n\in\mathbb{N}$ denotes the numbers of mode, $\lambda_{1(2)}$ is the symplectic eigenvalues of $\hat{\rho}_{1(2)}$,
\begin{equation}
\begin{aligned}
    &\bold{V}_{1(2)}\left[\nu\right]=\bold{S}_{1(2)}\left\{\bigoplus_{j=1}^n\left[\frac{\left(\lambda_{1(2),j}+1\right)^\nu-\left(\lambda_{1(2),j}-1\right)^\nu}{\left(\lambda_{1(2),j}+1\right)^\nu+\left(\lambda_{1(2),j}-1\right)^\nu}\otimes \bold{I}_2\right]\right\}\bold{S}_{1(2)}^T,\\
    &\bold{S}_{1(2)}\left\{\bigoplus_{j=1}^n\begin{pmatrix}
    0&1\\
    -1&0
    \end{pmatrix}
    \right\}\bold{S}_{1(2)}^T=\bigoplus_{j=1}^n\begin{pmatrix}
    0&1\\
    -1&0
    \end{pmatrix},\;\;\bold{C_{1(2)}}=\bold{S}_{1(2)}\left\{\bigoplus_{j=1}^n\lambda_{1(2),j} \bold{I}_2\right\}\bold{S}_{1(2)}^T,
\end{aligned}
\end{equation}
$\delta\bold{v}=\langle\bold{v}_1\rangle-\langle\bold{v}_2\rangle$, $\langle\bold{v}_{1(2)}\rangle$ is the quadrature mean and $\bold{C}_{1(2)}$ is the CM of $\hat{\rho}_{1(2)}$. 

Comparing with the QI performances under three idler distribution channels: ideal, direct one-way and $m$-relay QEC teleportation, we have the CMs of \emph{hypothesis 1} as $\left\{\bold{C}^{(1)}_{\text{ideal}},\bold{C}^{(1)}_{D},\bold{C}^{(1)}_{QT}\right\}$,
\begin{equation}
\begin{aligned}
    \bold{C}^{(1)}_{\text{ideal}}&=\begin{pmatrix}
        \left(2N_B+2\kappa N_S+1\right)\bold{I}_2&2\sqrt{\kappa}C_0\bold{Z}_2\\
        2\sqrt{\kappa}C_0\bold{Z}_2&\left(2N_S+1\right)\bold{I}_2
    \end{pmatrix},\;\;
    \bold{C}^{(1)}_D=\begin{pmatrix}
        \left(2N_B+2\kappa N_S+1\right)\bold{I}_2&2\sqrt{\eta\kappa}C_0\bold{Z}_2\\
        2\sqrt{\eta\kappa}C_0\bold{Z}_2&\left(2\eta N_S+1\right)\bold{I}_2
    \end{pmatrix},\\
    \bold{C}^{(1)}_{QT}&=\begin{pmatrix}
        \left(2N_B+2\kappa N_S+1\right)\bold{I}_2&2\sqrt{\kappa}C_0\bold{Z}_2\\
        2\sqrt{\kappa}C_0\bold{Z}_2&\left(2N_S+2\Sigma^2_{QT}+1\right)\bold{I}_2
    \end{pmatrix},
\end{aligned}
    \label{eq:cms}
\end{equation}
, and \emph{hypothesis 2} as $\left\{\bold{C}^{(2)}_{\text{ideal}},\bold{C}^{(2)}_{D},\bold{C}^{(2)}_{QT}\right\}$
\begin{equation}
\begin{aligned}
    \bold{C}^{(2)}_{\text{ideal}}&=\begin{pmatrix}
        \left(2N_B+1\right)\bold{I}_2&\bold{0}_2\\
        \bold{0}_2&\left(2N_S+1\right)\bold{I}_2
    \end{pmatrix},\;\;
    \bold{C}^{(2)}_D=\begin{pmatrix}
        \left(2N_B+1\right)\bold{I}_2&\bold{0}_2\\
        \bold{0}_2&\left(2\eta N_S+1\right)\bold{I}_2
    \end{pmatrix},\\
    \bold{C}^{(2)}_{QT}&=\begin{pmatrix}
        \left(2N_B+1\right)\bold{I}_2&\bold{0}_2\\
        \bold{0}_2&\left(2N_S+2\Sigma^2_{QT}+1\right)\bold{I}_2
    \end{pmatrix},
\end{aligned}
    \label{eq:cms2}
\end{equation}
where $\bold{0}_2$ is the $2\times2$ zero matrix. Calculating the symplectic eigenvalues of the CMs in Eq.~\ref{eq:cms} and Eq.~\ref{eq:cms2}, we substitute them into Eq.~\ref{eq:QCBformula} and numerically calculate the QCBs in Fig.~\ref{fig:QI_scheme}.

\subsection{CV quantum key distribution}
In the CV-QKD scheme, Alice and Bob preshared a TMSV state with CM
\begin{equation}
    \begin{pmatrix}
    V\bold{I}_2&\sqrt{V^2-1}\bold{Z}_2\\
    \sqrt{V^2-1}\bold{Z}_2&V\bold{I}_2
    \end{pmatrix},
\end{equation}
and have the mutual information at the limit of $V\gg1$
\begin{equation}
    \mathcal{I}_{AB}\approx\frac{1}{2}\log_2\left[\frac{V}{1+\epsilon}\right],
\end{equation}
where $V$ is the variance of the observed thermal mode if the state in Alice is traced out, $\epsilon$ is the variance of overall additive excess noise. Presumably, Eve adopts Gaussian attack, shown to be optimal among all collective attacks [103,104]), to the system. In this attack, the maximal accessible information is limited by the Holevo information,
\begin{equation}
    \chi_{BE}=S\left[\hat{\rho}_E\right]-\int p\left[x_B\right]S\left[\hat{\rho}^{x_B}_{E}\right]dx_B,
    \label{eq:Eve}
\end{equation}
 where $p\left[x_B\right]$ is the probability density function of Bob's measurement outcome $x_B$, $\hat{\rho}_{E}^{x_B}$ (or $\hat{\rho}_{E}$) are the density operators conditioned (or unconditioned) on Bob's result. Eq.~\ref{eq:Eve} can be derived as
\begin{equation}
    \chi_{BE}\approx\frac{1}{2}\log_2\left[e^2V\epsilon/4\right]
\end{equation}
and we obtain Eq.~\ref{eq:securitykeyrate} with the definition of SKR, $R\equiv\mathcal{I}_{AB}-\chi_{BE}$ (see more details in Ref.~\cite{Lodewyck07}).


\end{document}